\documentclass[sigconf,nonacm]{acmart}
\AtBeginDocument{%
  \providecommand\BibTeX{{%
    \normalfont B\kern-0.5em{\scshape i\kern-0.25em b}\kern-0.8em\TeX}}}

\copyrightyear{2024}
\acmYear{2024}
\acmVolume{}
\acmNumber{}

\acmConference[UIST '24]{The 37th Annual ACM Symposium on User Interface Software and Technology}{October 13--16, 2024}{Pittsburgh, PA, USA}
\acmBooktitle{The 37th Annual ACM Symposium on User Interface Software and Technology (UIST '24), October 13--16, 2024, Pittsburgh, PA, USA}
\acmDOI{10.1145/3654777.3676409}
\acmISBN{979-8-4007-0628-8/24/10}

\usepackage{color}
\usepackage{siunitx}
\usepackage{subcaption}
\usepackage{hhline}
\usepackage{tabularx}
\newcolumntype{Y}{>{\centering\arraybackslash}X}

\newcommand{\revise}[1]{#1}

\hypersetup{pdfauthor={Takeru Hashimoto, Yutaro Hirao}}
\begin{document}

\title[Selfrionette]{Selfrionette: A Fingertip Force-Input Controller for Continuous Full-Body Avatar Manipulation and Diverse Haptic Interactions}
\titlenote{Author-prepared version of the paper published in UIST '24, Article 137, pp. 1--14 (2024). The version of record is available at \url{https://doi.org/10.1145/3654777.3676409}. Copyright 2024 held by the owner/author(s).}

\author{Takeru Hashimoto}
\authornote{Both authors contributed equally to this research.}
\email{hashimoto@cyber.t.u-tokyo.ac.jp}
\orcid{0000-0003-4172-5692}
\affiliation{%
  \institution{The University of Tokyo}
  \city{Bunkyo}
  \state{Tokyo}
  \country{Japan}
}

\author{Yutaro Hirao}
\authornotemark[1]
\email{yutarohirao@gmail.com}
\orcid{0000-0003-3546-3454}
\affiliation{%
  \institution{Nara Institute of Science and Technology}
  \city{Ikoma}
  \state{Nara}
  \country{Japan}
}

\renewcommand{\shortauthors}{Hashimoto and Hirao}
\begin{abstract}
We propose Selfrionette, a controller that uses fingertip force input to drive avatar movements in virtual reality (VR). 
This system enables users to interact with virtual objects and walk in VR using only fingertip force, overcoming physical and spatial constraints. Additionally, by fixing users' fingers, it provides users with counterforces equivalent to the applied force, allowing for diverse and wide dynamic range haptic feedback by adjusting the relationship between force input and virtual movement.
\revise{To evaluate the effectiveness of the proposed method, this paper focuses on hand interaction as a first step.}
In User Study 1, we measured usability and embodiment during reaching tasks under Selfrionette, body tracking, and finger tracking conditions.
In User Study 2, we investigated whether users could perceive haptic properties such as weight, friction, and compliance under the same conditions as User Study 1.
Selfrionette was found to be comparable to body tracking in realism of haptic interaction, enabling embodied avatar experiences even in limited spatial conditions.
\end{abstract}

\ccsdesc[500]{Hardware~Emerging interfaces}
\ccsdesc[500]{Human-centered computing~Virtual reality}
\ccsdesc[300]{Human-centered computing~Accessibility technologies}
\ccsdesc[300]{Human-centered computing~Haptic devices}
\keywords{avatar manipulation, embodiment, haptics, force-based input}
\renewcommand{\shortauthors}{Hashimoto and Hirao}

\begin{teaserfigure}
  \includegraphics[width=\textwidth]{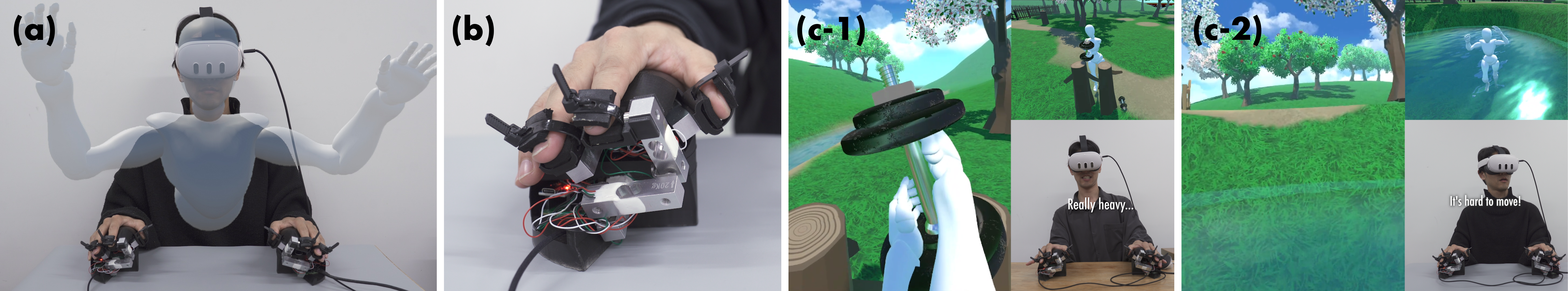}
  \caption{(a) Concept of Selfrionette revolves around enabling users to control the entire avatar's body through the force input of their fingers. (b) Selfrionette  Hardware: Comprised of multiple loadcells, Selfrionette offers a total of five degrees of freedom in force: three degree of freedom for the index finger and one degrees of freedom each for the thumb and middle finger. (c-1) Demonstration Scene 1: Users can move a virtual hand and grasp objects. The force required to move the arm varies depending on the weight of the object being grasped. (c-2) Demonstration Scene 2: Users can freely navigate through the virtual space. The force input required for walking also varies depending on changes in environmental characteristics, such as walking underwater.}
    \Description{(a) Concept of Selfrionette revolves around enabling users to control the entire avatar's body through the force input of their fingers. (b) Selfrionette  Hardware: Comprised of multiple loadcells, Selfrionette offers a total of five degrees of freedom in force: three degree of freedom for the index finger and one degrees of freedom each for the thumb and middle finger. (c-1) Demonstration Scene 1: Users can move a virtual hand and grasp objects. The force required to move the arm varies depending on the weight of the object being grasped. (c-2) Demonstration Scene 2: Users can freely navigate through the virtual space. The force input required for walking also varies depending on changes in environmental characteristics, such as walking underwater.}
  \label{fig:teaser}
\end{teaserfigure}
\maketitle

\hypersetup{
  pdftitle={Selfrionette: A Fingertip Force-Input Controller for Continuous Full-Body Avatar Manipulation and Diverse Haptic Interactions},
  pdfauthor={Takeru Hashimoto, Yutaro Hirao},
  pdfsubject={Author-prepared version; UIST 2024, Article 137, pages 1--14; DOI: 10.1145/3654777.3676409},
  pdfdoi={10.1145/3654777.3676409},
  pdfcopyright={Copyright 2024 held by the owner/author(s).},
  pdfpublication={Proceedings of the 37th Annual ACM Symposium on User Interface Software and Technology (UIST '24)},
  pdfpagerange={1-14},
  pdfurl={https://doi.org/10.1145/3654777.3676409}
}

\section{Introduction}
Can people liberate themselves from their physical and spatial constraints?
Virtual reality (VR) technology has emerged as a means of interfacing with various entities through virtual bodies without the need for a physical presence.
A virtual body, an avatar, functions as an altered body capable of interacting with the virtual world independently of the physical body. 
Information obtained by moving an avatar is analogous to information gained from moving one's own body.
\revise{The ideal state when using an avatar is achieved when predictions align through a continuous loop of perception and action \cite{friston2010action}. 
In this state, the avatar becomes embodied, naturally and unconsciously performing actions based on the information it receives, as if it were an extension of the user's own body.
}

Typical avatar movement relies on motion tracking, which mirrors real-world movements in VR. While this method is intuitive, it may not be fully effective for users with spatial limitations or physical disabilities.
\revise{
Moreover, using motion-tracking methods makes it difficult to present haptic information when interacting with virtual objects. 
The lack of haptic information reduces  the sense of embodiment with the avatar.
Although various haptic devices have been proposed, they often require large equipment to replicate real-world physical phenomena.
}
In a 1965 article, “The Ultimate Display,” Ivan Sutherland suggested that ``there is no reason why the objects displayed by a computer have to follow the ordinary rules of physical reality. \cite{Sutherland1965-jy}'' 
To address these issues, we move beyond direct movement mappings and explore more flexible interactions.

\revise{
Analyzing existing avatar control methods, we identify two essential components for enabling whole-body movements: (1) the type of user movement input, which can be either motion-based or force-based, and (2) the mapping of this input to the avatar's movements, which can be either direct or indirect mapping. 
Here, force-based input methods are known for conveying congruent haptic interactions essential for achieving embodiment \cite{hirao2023move}.
Moreover, indirect mapping cannot be constrained by spatial and physical limitations.
}
Prior work includes FingerMapper \cite{Tseng2023-cs}, which uses the user's finger movements to move the avatar's hands.
\revise{
This method is less spatially constrained by indirect mapping, but it is not easy to represent haptic interactions because it is motion-based.}
Imanaga et al. proposed a method that fixes the user's whole body and moves the avatar's whole body based on the force that tries to move the whole body \cite{Imanaga2022-qo}.
\revise{
This method is force-based and potentially excels in haptic expression (though this has yet to be verified in their paper), but as a direct-mapping method, it requires large equipment and full-body constraints to represent whole-body movements.}

Our study addresses haptic interaction issues through a force-based approach, enabling avatar manipulation under environmental and physical constraints through indirect mapping.
Therefore, we propose a system that converts force input from the fingers into full-body movements of an avatar, naming it the \textit{"Selfrionette,"} which manipulates \textit{self} entire body like a \textit{marionette} controlled by the fingers. 
\revise{To evaluate the effectiveness of the proposed method in experiments, this paper focuses on hand interaction as a first step, although we implemented walking interactions on trial to discuss the proposed method's potential and possible future directions.}

The contributions of this paper can be summarized as follows:
\begin{enumerate}
    \item Development of a system for converting force from finger input into avatar movements and mapping.
    \item Investigation into how different body manipulation methods affect motor skill and sense of embodiment \revise{especially in hand interaction}.
    \item Investigating the quality of haptic interactions and sense of embodiment in different body manipulation methods \revise{during hand interaction}.
\end{enumerate}

\section{Related work}

\subsection{Avatar manipulation without morphological correspondence}
\subsubsection{Motion-based manipulation}
The first option for the input system would be motion-based manipulation to manipulate a virtual body in VR or digital puppetry.
Of motion-based techniques, the most conventional approach is full-body motion tracking, where the user's full-body movements in the real world are mapped to those of the avatar in VR.
Several papers endeavor to convert these full-body motion inputs even into virtual movements without morphological correspondences by extracting features from the motion data (e.g., \cite{yamane2010animating, chen2012kinetre, kim2022humanconquad, egan2023neurodog, li2023ace, reda2023physics}). The method utilizing full-body movements is intuitive for users and has a high degree of freedom of input. However, it requires adequate space to move and has difficulties for users with physical disabilities.

Another trend in manipulating virtual avatars with motion-based techniques is using hand gestures with indirect mappings. Some studies construct a continuous mapping of joints between hands and avatars (e.g., \cite{komura2006real, cheng2017create, jiang2023handavatar, momeni2015dranimate, tseng2023fingermapper, kondo2020re}). Other papers create a mapping rule referring to a marionette system (e.g., \cite{leite2017mani, oshita2013character}). Moreover, gestures to activate specific actions allow users to make complex motions (e.g., \cite{farahanipad2020hand, velho2020expanded, liang2017hand, sheikholeslami2017cooperative}). These approaches wisely utilize the dexterity of human hands and enable relatively complex virtual actions with intuitive inputs. However, most studies focus on avatar posture control and do not investigate haptic interaction with virtual objects well. Ujitoko and Hirota proposed a method of presenting tactile feedback on a virtual floor while walking in VR with finger motions \cite{ujitoko2014interpretation}. Furthermore, Xiang et al. propose a method of presenting geometric properties of virtual ground and contact sensation with virtual objects while virtual foot interactions with finger motions \cite{wang2020miniature}.
However, to the best of our knowledge, no papers endeavor to realize more various haptic interactions, such as weight, compliance, and friction, with such a morphologically inconsistent avatar manipulation system.

Another method is to use a physical prop. It can be a motion controller (e.g., \cite{laszlo2000interactive, dontcheva2003layered, oore2002desktop}) or a physical puppet that has multiple joints (e.g., \cite{numaguchi2011puppet, glauser2016rig, saint2023tangible}). These approaches are also intuitive, especially regarding posture control, because the physical prop corresponds to the digital puppet. However, they are instead a tool than an embodied virtual self, and these methods do not achieve haptic interactions with virtual objects.

\subsubsection{Force-based manipulation}
Some papers endeavor to convert force-input into virtual motions.
"Tsumori" control proposed by Niwa et al. involves operating an avatar (robot) by associating the user's force input patterns with pre-segmented action units \cite{niwa2010detection}. This technique relaxes the requirement for spatiotemporal congruence between input and output to the level of discrete spatiotemporal resolutions, termed behavioral segments. The system allows for inferring user control intentions from low-degree-of-freedom inputs and enables high-degree-of-freedom operations. However, the degree of freedom in avatar movements is possible through "Tsumori" control, and the learning cost required for the necessary associations is in a trade-off relationship. Additionally, compared to continuous input-output mapping, "Tsumori" control has inherent limitations to the operational degrees of freedom while interacting with virtual objects.
Moreover, there are challenges in applying the visual-based haptics technique, which dynamically alters the correspondence between force input and motion.

Imanaga et al. proposed Motion-lesss VR where the user's body is fixed to a device equipped with force sensors \cite{Imanaga2022-qo}. By attempting physical movements in this state, the motion intent is calculated inversely from the torque obtained from each force sensor and reflected as the movement of a virtual body.
However, their method requires restraining the entire physical body to control the entire virtual body, which may pose challenges to usability in terms of fatigue and comfort.
Moreover, when full-body restraint is employed, the user cannot move, presenting potential safety issues in emergencies or malfunctions.
Additionally, the mechanism must be capable of automatic attachment and detachment. 
To sense the user's motion intent as torque, force sensors must be calibrated and positioned considering individual anatomical differences, complicating and increasing the cost of the system.

\revise{
\subsection{Embodiment in Virtual Avatar}
Sense of embodiment (SoE)--the recognition that "this is my body"-- plays an essential role in sense of immersion/presence, i.e., sense of being there, in VR \cite{slater2010simulating}.
Although the mechanism underlying SoE has yet to be fully elucidated, many papers endeavor to investigate its key components.
SoE in the context of VR research consists of three factors: the sense of agency (SoA), sense of body ownership (SoBO), and sense of self-location (SoSL) \cite{kilteni2012sense}. 
SoA refers to a feeling of causing actions or their consequences \cite{haggard2012sense}, which arises from sensory attenuation caused by the contrast between predicted bodily movement outcomes (efference copy) and actual sensory input \cite{adams2013computational}; SoBO refers to "the sense that one's own body is the source of sensations"  \cite{tsakiris2006having}, which results from Bayesian multisensory integration of sensory predictions and sensory inputs related to the self-body \cite{kilteni2012sense}; and SoSL "refers to one's spatial experience of being inside a body" \cite{kilteni2012sense}. To enhance SoA in VR, it is essential to map user's motion (input) to the virtual body's motion (output) in real-time or near real-time mapping; to enhance SoBO in VR (by bottom-up influences), it is essential to increase the sensory correlations between physical and virtual body such as visuo-tactile and visuo-proprioceptive correlations, for example; and to enhance SoSL, it is fundamental to have first-person perspective \cite{kilteni2012sense}.
We designed Selfrionette to meet the essential requirements of these SoE. Detailed descriptions are provided in Section \ref{sec:rationale}.
}

\subsection{Visual based haptics}
The paper utilizes a visual-based haptics approach to achieve haptic interaction in morphologically inconsistent manipulation systems.
Conventional studies of visual-based haptics are classified under "Pseudo-haptics techniques," which modify haptic perception by appropriately adjusting visual feedback in response to user inputs \cite{Pusch2011, Ujitoko2021}. Since the first paper on pseudo-haptics in 2000 \cite{Lecuyer2000}, the pseudo-haptics techniques have been studied for various haptic perceptions such as texture (e.g., \cite{Narumi2017-yk, ujitoko2019presenting, lecuyer2004feeling}), compliance (e.g. \cite{Lecuyer2000, punpongsanon2015softar, Argelaguet2013, kumar2017mechanics, tatezono2009effect}), and weight (e.g. \cite{taima2014controlling, rietzler2018breaking, Samad2019-va}).
In this manner, pseudo-haptics techniques have the advantage of expressing various haptic perceptions primarily through visual feedback without the need for complex devices. 
However, when used with the currently mainstream motion-input systems in VR, a discrepancy arises between the user's actual and visual movements in VR. 
This discrepancy is problematic as it can lead to discomfort in interactions and a reduction in the sense of immersion \cite{Pusch2011, rietzler2018breaking, honda2013imposed}. 
To address this issue, recent research has indicated that force-input systems are more effective than motion-input systems in terms of the efficacy of pseudo-haptics techniques.
Indeed, psychophysical experiments suggest that the pseudo-haptics techniques can alter weight perception only by 3--5\% with motion-input system \cite{taima2014controlling, Samad2019-va}, but can modify compliance perception by 20--35\% with force-input (or isometric) systems \cite{lecuyer2001boundary, tatezono2009effect}. 
Moreover, in recent investigations comparing pseudo-weight through motion-input versus force-input system, it has been demonstrated that the force-input system exhibits superior pseudo-haptics expressiveness \cite{hirao2023move}. 
Building on the aforementioned findings, this study determined to integrate the element of force-input into an avatar manipulation system.

\section{Selfrionette} 
\subsection{Rationale behind Selfrionette}
\label{sec:rationale}
This section discusses the rationale behind Selfrionette as a method for embodied virtual body manipulation that can overcome physical and spatial constraints and enhance virtual haptics presentation. 
Selfrionette is effective against these constraints by controlling a full-body virtual avatar with only fingers. 
The critical inquiry is whether such a system, with morphologically distinct input-output correspondences, can update the body schema and elicit a sense of embodiment.
Research in the field of psychology regarding the sense of embodiment in avatars and objects has demonstrated that the correlation between different sensory modalities is crucial for the sense of body ownership. 
Specifically, visual information must match tactile and proprioceptive information for perception to occur accurately \cite{kilteni2012sense}. 
Additionally, in the domain of prosthetics research, it has been suggested that even if the locations of visual and tactile stimuli differ, users can perceive the stimuli as consistent information as long as the expected and actual tactile stimuli types match (modality match) \cite{svensson2017review}.
This modality-matched feedback is known to reduce the cognitive effort required for users to adapt to prosthetic operation compared to other sensory substitution methods \cite{antfolk2013sensory}, thereby enhancing the sense of embodiment \cite{marasco2011robotic}.
In this regard, Selfrionette employs force input, where the user-exerted force on a sensor triggers an equivalent reactive force, altering skin deformation and proprioceptive information. 
The reactive force and the corresponding feedback vary as the input force varies. 
Thus, Selfrionette potentially offers a wide dynamic range of modality-matched feedback during VR interactions.

Furthermore, manipulating virtual bodies through force input allows for haptic interactions by varying the required input force according to the parameters of the virtual objects and environments. 
Indeed, Hirao et al. demonstrated that weight perception during force input manipulation could be presented with a wider perceptual intensity range than conventional pseudo-haptic technique using motion tracking \cite{hirao2023move}. 
This suggests that the visual information of the virtual hand and the haptic information from the user's fingers were perceived as originating from the same source (sensory correlation was established), leading to integrated and enhanced haptic perception.

Based on these considerations, Selfrionette holds the potential to enable embodied virtual body manipulation that overcomes physical and spatial constraints and extends the capabilities of virtual haptics presentation.

\subsection{Hardware}
\begin{figure}
    \centering
    \includegraphics[width=\linewidth]{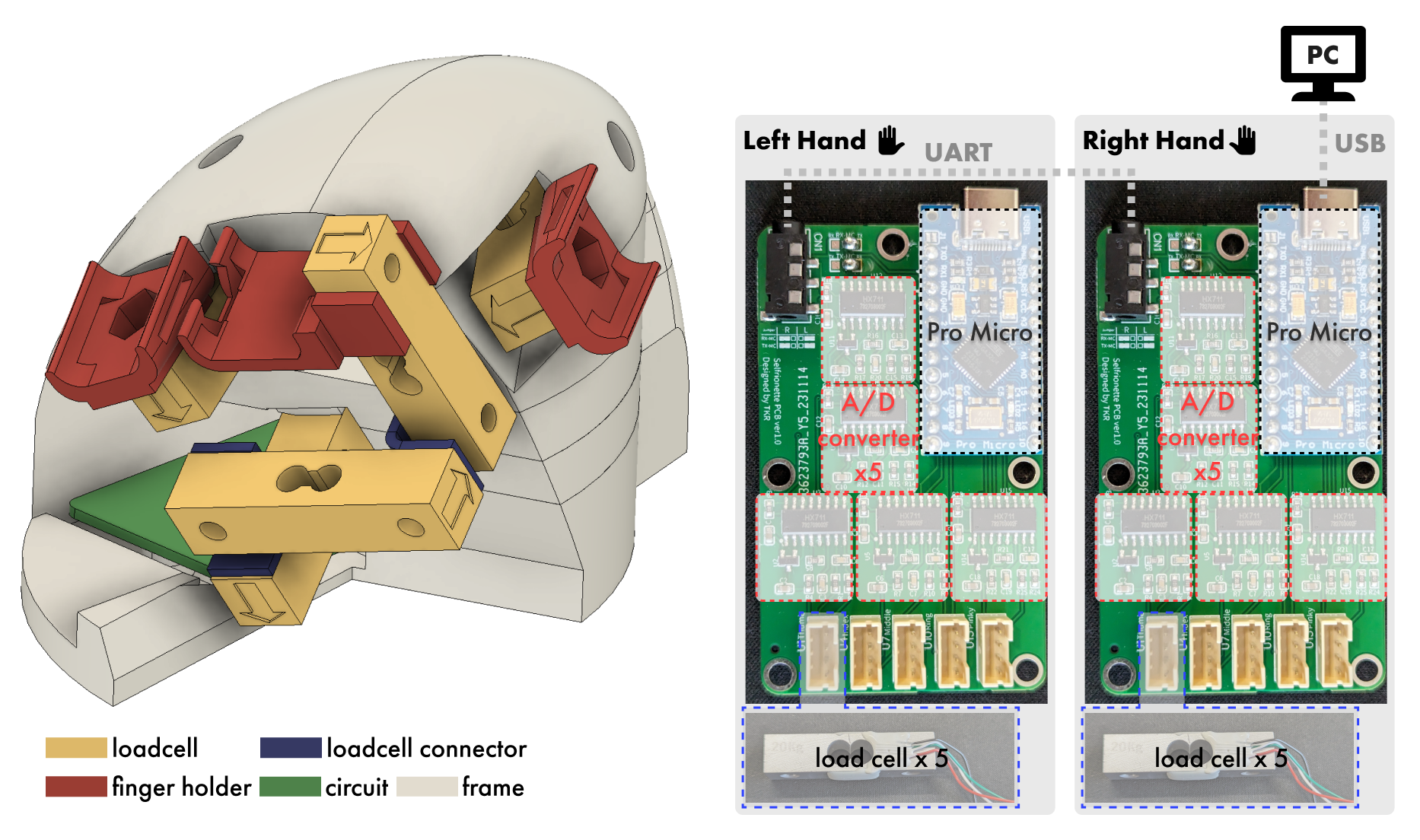}
    \caption{Left) Components.  Right) System configuration of Selfrionette.}
    \Description{Left) Components.  Right) System configuration of Selfrionette.}
    \label{fig:Selfrionette-hard}
\end{figure}

Figure \ref{fig:Selfrionette-hard} shows the appearance and components of the device, as well as the system diagram.
The hardware for one hand consists of five single-beam load cells and a spherical casing, which allows users to place their hands in a natural posture.
Each load cell measures \SI{12.5}{} $\times$ \SI{12.5}{} $\times$ \SI{55}{mm} and can convert a force input of up to 20kg into an analog signal. 
The analog signals from the load cells are converted into digital signals by an HX711 chip at \SI{80}{Hz} and read by a microcontroller (Pro Micro).

The five load cells are allocated, with one for the thumb, one for the middle finger, and three for the index finger in an orthogonal arrangement. 
This arrangement allows for the acquisition of input forces in one degree of freedom for the pressing direction of the thumb and middle finger and three degrees of freedom for the index finger in the up, down, forward, backward, left, and right directions.
Reusable cable ties are also used to secure the user's fingers to the finger holder. 
Inserting low-resilience urethane foam between the fingers and the cable ties allows for comfortable and secure finger fixation.
By connecting two devices, ten degrees of freedom input can be acquired for both hands.
\subsection{Mapping from force to avatar motion}
\begin{figure}
    \centering
    \includegraphics[width=\linewidth]{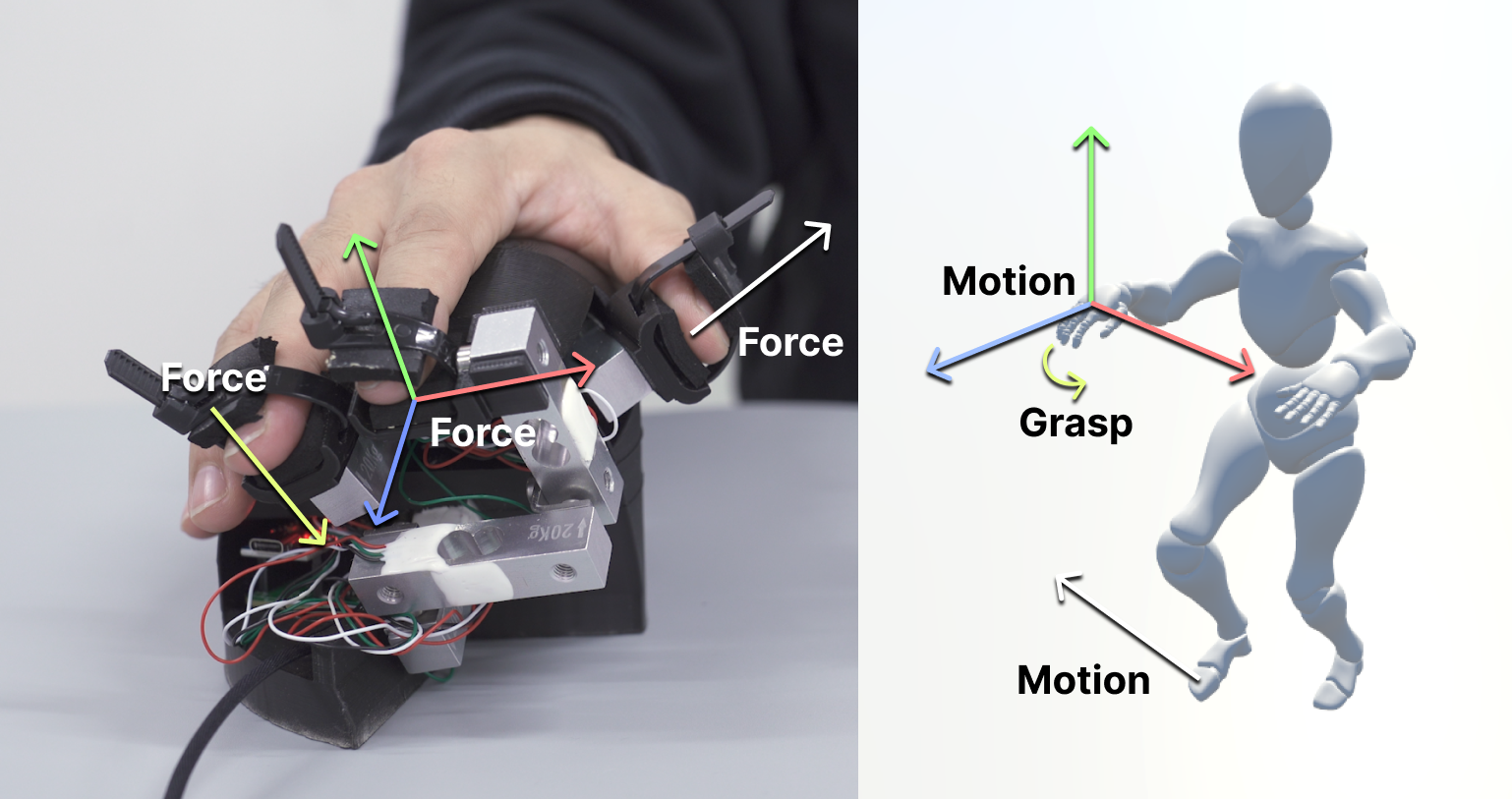}
    \caption{Correspondence between force inputs and avatar movements.}
    \Description{Correspondence between force inputs and avatar movements.}
    \label{fig:correspondence}
\end{figure}

Next, the conversion from force input to motion of a virtual avatar is explained. 
Figure \ref{fig:correspondence} shows the correspondence between force inputs from the device and avatar movements.
The avatar's posture is updated using inverse kinematics (IK) with FinalIK\footnote{http://root-motion.com/}, setting target points for the head, hands, and feet.

The target point for the head is configured to follow the position of the Head-Mounted Display (HMD).
\revise{
The mapping between fingers and virtual movements was based on the following principles.

\noindent\textbf{Hand Motion:} To move the hand in three-dimensional space, at least three degrees of freedom are required. 
Each fingertip possesses at least three degrees of freedom, making them suitable for controlling the hands of an avatar. 
Specifically, the index finger, used for general pointing, is optimal for the avatar's hand to point at targets or reach destinations.
We mapped the three degrees of freedom of the index finger to the hand's movement.

\noindent\textbf{Foot Motion:} Walking requires at least two degrees of freedom. 
Humans can move their arms and legs independently, so we assigned both thumbs, which can operate independently of the index fingers, to control the legs' movement.

\noindent\textbf{Grasping:} The middle finger was chosen to control the hand's grasping movement because it is the easiest to control among the remaining fingers.
}
For moving the legs and hands, the force applied to each load cell is converted into a force acting on the corresponding target point (whose mass $m_v$ is virtually 1kg) in VR space.
The position of target point is then updated, and the updated target point is followed by the hands or feet.

The relationship between the force applied to the load cell $F_l$ and the force applied to the target in the VR space $F_v$ can be freely designed.
In this study, stable motion was achieved by defining the behavior as shown in Equation \ref{eq:relation}.
Specifically, the force applied to the load cell is proportional to the force applied to the target by the proportionality constant $\alpha$, and the behavior of the motion of each target $\ddot{x}$ is described by a spring-damper model.

\revise{
\vspace{2pt}
\noindent\textbf{Equation for Avatar Motion Generation}
\begin{equation}
\begin{array}{@{}ll@{}}
    F_v = \alpha F_l & \; \text{\small convert fingertip force to virtual force} \\
    m_v \ddot{x} + c \dot{x} + k (x - x_0 ) = F_v & \; \text{\small move the avatar's target position}
\end{array}
\label{eq:relation}
\end{equation}
}

In this study, the spring constant $k$ and damping coefficient $c$ were both set to \SI{10}{N/m}, \SI{10}{Ns/m} respectively. 
The origin of the spring-damper system $x_0$ is positioned 0.375m in front of each shoulder for the hands and at the central position of each foot when spread shoulder-width apart (the avatar's initial stance) for the feet. 
The force acting on the target point for the feet is presented along an axis angled 45 degrees upward and forward from the spring-damper system's origin.

Walking in VR is enabled through \revise{thumbs alternating inputs to achieve} virtual on-the-spot stepping with the SHeF-WIP (Stepping Height Frequency-Walking In Place) method \cite{hirao2022revisiting}, which takes into account both the stepping frequency and height to simulate natural walking. 
Directional changes and backward movements are achieved by adjusting the stepping height between the left and right feet.

The degree of hand grip is controlled by the middle finger, with a continuous range from 0 (hand open) to 1 (hand clenched). 
When a virtual object is near the hand, the system detects if the hand clenches beyond a certain threshold, enabling the object to be grasped.
The AutoHand\footnote{https://assetstore.unity.com/packages/tools/game-toolkits/auto-hand-vr-interaction-165323}  tool is used to automatically calculate the gripping shape of the hand according to the shape of the virtual object, ensuring a natural grasping hand shape.

Consequently, this version of the Selfrionette allows for the control of 3DoF positions of the hands, the vertical position of the feet, and interactions such as grasping/releasing objects and walking. 

\section{Haptic Interaction via Selfrionette System}\label{sec:hap}

\begin{figure*}
    \centering
    \includegraphics[width=\linewidth]{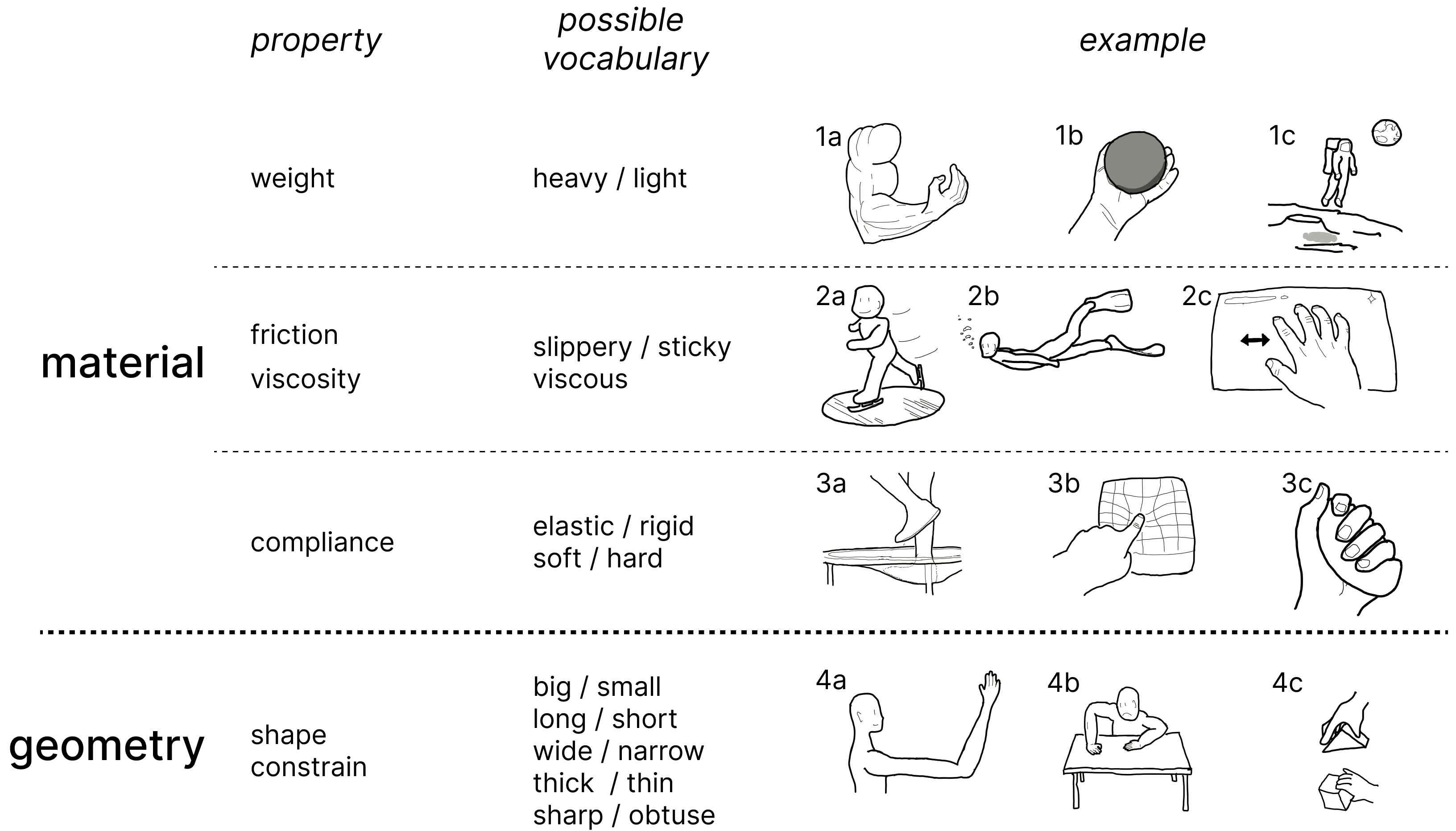}
    \caption{Haptic Interaction Space archived by Selfrionette system.}
    \Description{Haptic Interaction Space archived by Selfrionette system.}
    \label{fig:haptic-interaction}
\end{figure*}
This chapter discusses the methods for delivering haptic feedback using the Selfrionette system and categorizes the haptic properties that can be simulated. 
To categorize haptic properties, We draw upon studies in motion-based pseudo-haptics \cite{Lecuyer2009-cl, Ujitoko2021-zb}, redirection \cite{ban2012modifyingsurface}, impedance control in haptic rendering \cite{10.1145/3526113.3545650, Hashimoto2023-dw, 10.1145/3588037.3595390}, as well as categorization frameworks proposed by Jones and Lederman \cite{Jones2006-ln} and Okamoto et al. \cite{Okamoto2013-sh} to organize these properties.
The classification of haptic properties revolves around two main aspects: material and geometric properties.
Material properties encompass texture, compliance, and weight. 
Texture is further broken down into five main components: fine roughness, macro roughness, friction, hardness, and warmth.
Geometric properties, on the other hand, include curvature, orientation, and size.
Selfrionette operates by translating fingertip force input into motion within the virtual space. While force input primarily correlates with proprioception, tactile properties such as fine roughness and warmth, mainly perceived through fingertip receptors, cannot be stimulated in this study.
Figure \ref{fig:haptic-interaction} illustrates the classification of haptic interactions that can be expressed with the Selfrionette system. 
Multiple examples are provided for each property, along with associated vocabulary.
A detailed breakdown of each property and its implementation approach follows.

\subsection{Geometric Properties}
Geometric properties are directly tied to the positional information of the body. 
By leveraging reaction forces occurring upon contact between the body and objects, Selfrionette can constrain the avatar's position to reflect the geometric properties of the object's surface. 
Specifically, at the contact point, the force $f$ applied by the body to the object is decomposed into the normal direction force $f_n$ and the parallel direction force $f_p$. 
The parallel direction force $f_p$ is used to update the contact point, while the normal direction force $f_n$ is ignored. 
As we discuss in the \textit{material properties} section, this normal force $f_n$ is used to represent compliance of the object.
This enables the representation of various geometric attributes, such as environmental surface constraints or shape constraints of objects.

For example, it can represent the plane constraint of hands in the environment (4b) or the constraint by exact shape when grasping an object with the fingers (4c).
It is also possible to change the geometric properties of the body itself (4a) since there are no spatial constraints on positional control by force input.

\subsection{Material Properties}
Material properties are linked to positional, velocity, and acceleration information. 
Selfrionette can simulate weight, viscosity, and compliance.
These properties correspond to the body's acceleration, velocity, and displacement during interaction.
Diverse haptic interactions are realized by adding force $f_m$ to the body motion equation shown in Eq (\ref{eq:relation})
\revise{
\begin{equation}
    \underbrace{m_v \ddot{x} + c\dot{x} + k(x - x_0)= F_v}_{\mbox{\small equation (\ref{eq:relation}) for avatar motion}} + \underbrace{f_m}_{\mbox{\small material rendering force}}
    \label{eq:haptic_interaction}
\end{equation}
}
\subsubsection{Weight}
The mass component in impedance control can represent weight. (1b) To perceive the weight of an object, unsupported grasping is necessary, requiring acceleration and gravity-based feedback corresponding to the object's weight when grasped.
For implementation, with the object's position denoted as $x$, gravity vector as $g$, object mass as $m$, it can be expressed as:
\begin{equation}
    f_m = -m \ddot{x} - mg
    \label{eq:weight}
\end{equation}
Moreover, it is possible to alter the sense of one's body weight, not just objects placed in the environment (1a).
Regardless of object grasping, the force required to move the arm itself can be adjusted according to Eq \ref{eq:weight}. 
Furthermore, it would be possible to change the magnitude of the gravity of the environment felt by the whole body. In this case, the gravity component of the physical phenomenon $g$ is changed (1c).

\subsubsection{Viscosity / Friction}

Viscosity describes the properties of a medium, not of an object or body. 
Depending on the properties of the fluid, it produces a resistance force proportional to its velocity. 
With viscous constant as $c$, viscous force can be expressed as:
\begin{equation}
    f_m = -c \dot{x}
    \label{eq:viscosity}
\end{equation}
This allows us to describe the perception of swimming in water (2b) or walking through mud (2a). 

Friction between object surfaces can also be represented. 
Friction can be static, in which there is no motion between two surfaces, or dynamic when motion occurs between two surfaces.
These are expressed using the vertical drag force $f_n$, the static friction coefficient $\mu$, and the dynamic friction coefficient $\mu'$.
\begin{equation}
    f_m = 
      \begin{cases}
    - f_p & \text{if $|f_p| \leq \mu |f_n|$ (dynamic friction)} \\
    - \mu' |f_n| \text{dir}(f_p) & \text{if $|f_p| > \mu |f_n|$(static friction)}
  \end{cases}
    \label{eq:friction}
\end{equation}
This additional friction could represent the sensation of tracing rough surfaces like rubber (2c) or moving over rough or smooth surfaces like ice fields (2a).
\subsubsection{Compliance}

Compliance refers to an object's elastic behavior and is simulated using the spring component in impedance control. 
When a part of the body contacts an object's surface, the system generates a virtual reaction force proportional to the displacement from the point of contact $x_c$, and simultaneously elastically deforms the object. 
\begin{equation}
    f_m = -k (x - x_c)
    \label{eq:compliance}
\end{equation}
This allows the user to represent the softness felt when touching (3b) or grasping (3c) an object or the elasticity felt with the feet, as in the case of a trampoline (3a). 

This classification elucidates the spectrum of haptic interactions achievable with Selfrionette.
In user study 2, we will delve into user perception of these haptic interactions' realism and their ability to discern material differences.

\section{User Study 1}\label{sec:u1}
We conducted a user study to investigate the extent to which Selfrionette enables bodily movements in VR. 
The purpose of the experiment is to compare the effectiveness of Selfrionette in achieving virtual motions with other conceivable methods. 

\subsection{Experiment Conditions}
As a comparative method, we exclude the whole-body force-based method among the four combinations of tracking and input methods introduced in Related Work, and compare three other methods. 
It should be noted that the whole-body force-based method was not chosen as it compromises the space-saving advantage of this study due to its bulky apparatus. 
Hereafter, we refer to the motion-based method using the entire body as the \textit{Body} condition, the motion-based method using fingers as the \textit{Finger} condition, and the force-based method using fingers as the \textit{Selfrionette} condition. 

\subsection{Participants}
A total of 18 individuals participated, consisting of 9 males and 9 females, with an average age of 25.83 $\pm $ 7.13 years old. 
All participants were recruited through social media, unaware of the true purpose of the experiment. 
All participants self-reported as right-handed. 
They signed an approved statement of consent and were compensated with an Amazon gift card worth approximately \$10.

\subsection{Materials}
\subsubsection{Apparatus}
The experimental setup included a Windows-based computer, an HMD (Meta Quest 3), and the Selfrionette device. 
The experimental program was developed using Unity 3D.

\subsubsection{Implementation of finger condition}
For the finger condition, we implemented a system similar to FingerMapper \cite{tseng2023fingermapper}, which translates the coordinates of the tip of the index finger into reference coordinates for the inverse kinematics (IK) of a virtual hand. The operational range of the virtual hand was confined to a hemisphere, centered at a point \SI{0.15}{m} in front of the avatar's shoulder and with its apex \SI{0.6}{m} forward from the shoulder. Here, it is known that the human index finger's range of motion includes 148 degrees for Flexion/Extension and 50 degrees for Abduction/Adduction \cite{palastanga2006anatomy}. Thus, in our implementation, we achieved a coordinate transformation that accounts for both the index finger's range of motion and the operational range of the virtual hand across each axis in three-dimensional space.

\subsection{Reaching Task}
\begin{figure*}
    \centering
    \includegraphics[width=\linewidth]{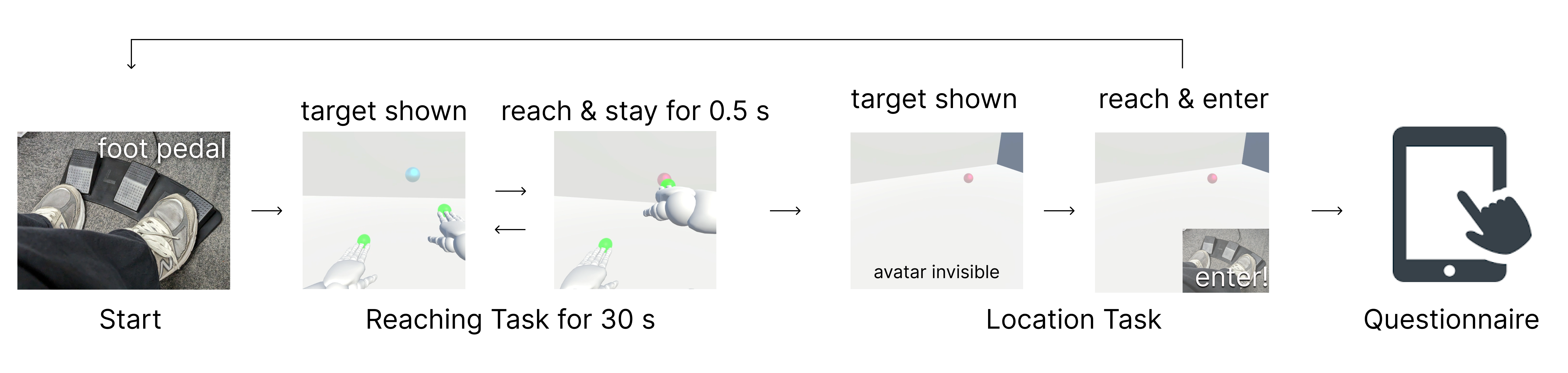}
    \caption{Experiment Procedure.}
    \Description{Experiment Procedure.}
    \label{fig:exp1_procedure}
\end{figure*}

Participants engaged in a task where they reached for sequentially appearing targets in the virtual space to assess the accuracy and speed of their bodily movements. 
Specifically, they were instructed to touch target spheres that appeared within the space using their avatar's arms. 
Figure \ref{fig:exp1_procedure} shows the procedure of reaching task.
This task began upon participants pressing a foot pedal situated at their feet.
Spheres, each with a diameter of \SI{5}{cm}, would randomly emerge in front of participants in a specific area (centered 1.5 m from the ground and 0.45 m in front of the avatar, with a width $\times$ height $\times$ depth of \SI{0.8}{} $\times$ \SI{0.3}{} $\times$ \SI{0.3}{m}). 
Participants moved their avatar's hand to approach these spheres, aiming to maintain contact for 0.5 seconds.
Additionally, if participants did not complete the reaching and contacting within 5 seconds, the task ended forcibly. 
Upon completion of a reaching action, a new target sphere appeared at a different location, initiating the next reaching task. 
These procedures were repeated for 30 seconds.
The number of successful tasks and the time taken for each task were recorded.

\subsection{Self localization task}
This task was designed to gauge participants' ability to develop a sense of self-location. 
Following the completion of a 30-second trial in the reaching task, participants adjusted their avatar to a predetermined position. 
Subsequently, participants experienced a condition where their avatars were not visible, prompting them to reach for randomly appearing target sphere within the space. 
Participants indicated their responses by pressing the foot pedal when they believed their body had made contact with the sphere.
The distance error between the position of the invisible avatar's hand and the position of a displayed target sphere was recorded.

\subsection{Questionnaire}
A questionnaire was administered to evaluate participants' subjective experiences regarding embodiment, system usability, and fatigue. 
Embodiment was assessed using Gonzalez-Franco and Peck's VR-specific avatar embodiment questionnaire \cite{Gonzalez-Franco2018-jq}. 
We omitted items that were not applicable to our study context, such as the question regarding haptic sensation.
Consequently, the questionnaire consisted of ten items and three subsets of questions: body ownership (Ownership), agency and motor control (Agency), and body location (Location) (Table \ref{tab:embodiment}).
Additionally, questions concerning fatigue were included to assess subjective fatigue levels associated with different manipulation methods. 
Each response was scored on a seven-point Likert scale (-3 = strongly disagree; 3 = strongly agree).
System usability was evaluated using the ten-item System Usability Scale \cite{Brooke1996-ku}. 
Furthermore, participants were asked to rank their subjective preferences for the three methods following each condition, providing reasons for their choices.

\begin{table}
    \centering
    \begin{tabularx}{\linewidth}{lX}
        \hline \noalign{\vspace{2pt}}
        Subscale & Question \\ \noalign{\vspace{2pt}}
        \hline \noalign{\vspace{2pt}}
        Ownership & 1) I felt as if the virtual hand (sphere) was my hand. \\
                & 2) It felt as if the virtual hand (sphere) I saw was someone else’s. \\
                & 3) It seemed as if I might have more than one hand.  \\[2pt]
        \hline \noalign{\vspace{2pt}}
        Agency & 4) It felt like I could control the virtual hand as if it was my own hand. \\
                & 5) The movements of the virtual hand were caused by my movements. \\
                & 6) I felt as if the movements of the virtual hand were influencing my own movements. \\
                & 7) I felt as if the virtual hand was moving by itself. \\[2pt]
        \hline \noalign{\vspace{2pt}}
        Location & 8) I felt as if my hand was located where I saw the virtual hand. \\
                 &   9) I felt out of my body. \\
                 &   10)  I felt as if my (real) hand were drifting toward the virtual hand or as if the virtual hand were drifting toward my (real) hand. \\[2pt]
        \hline \noalign{\vspace{2pt}}
        Fatigue & 11) This condition makes me exhausted. \\
        \hline \noalign{\vspace{2pt}}

    \end{tabularx}
    \caption{Questionnaire to Measure the Sense of Embodiment Based on \cite{Gonzalez-Franco2018-jq} and Fatigue.}
    \label{tab:embodiment}
\end{table}

\subsection{Procedure}
The experiment comprised three main components: the reaching task, the self-localization task, and a questionnaire.
Participants engaged in these activities while wearing the HMD. The entire procedure lasted approximately 45 minutes.
The participants
\begin{enumerate}
\renewcommand{\labelenumi}{\arabic{enumi})}
    \item read and signed an experimental consent form.
    \item received instructions and an overview of the task.
    \item repeated the reaching task (30 sec) and self-localization task ten times.
    \item completed the embodiment, fatigue, and SUS questionnaire.
    \item repeat (3) and (4) in each condition. 
    \item completed the post questionnaire.
\end{enumerate}
The order of experimental conditions was counterbalanced.

\subsection{Result}

\begin{figure}
    \centering
    \includegraphics[width=\linewidth]{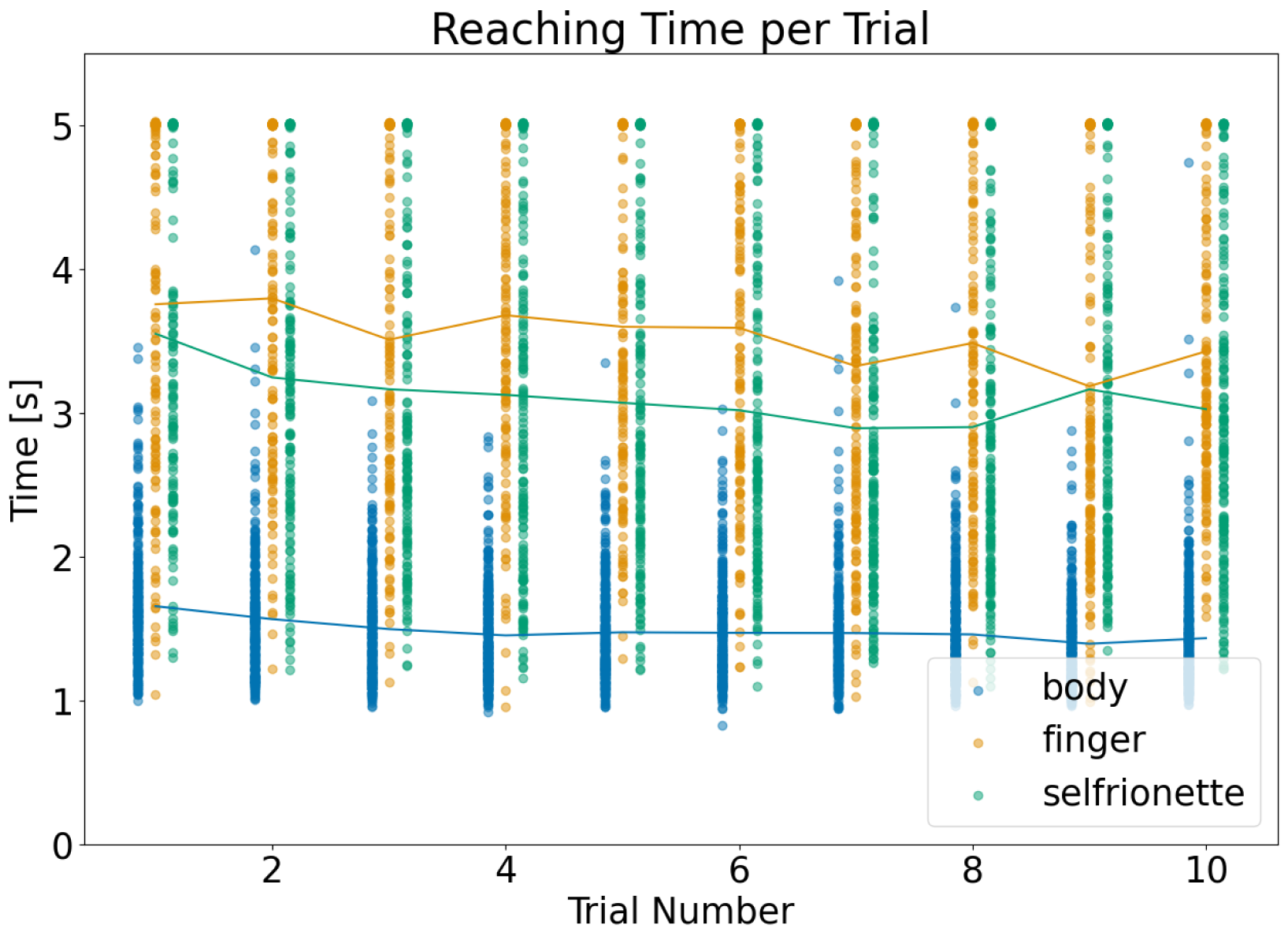}
    \caption{Transition of task completion time per trial}
    \Description{Transition of task completion time per trial}
    \label{fig:exp1_task_time}
\end{figure}

\begin{figure}
    \centering
    \includegraphics[width=\linewidth]{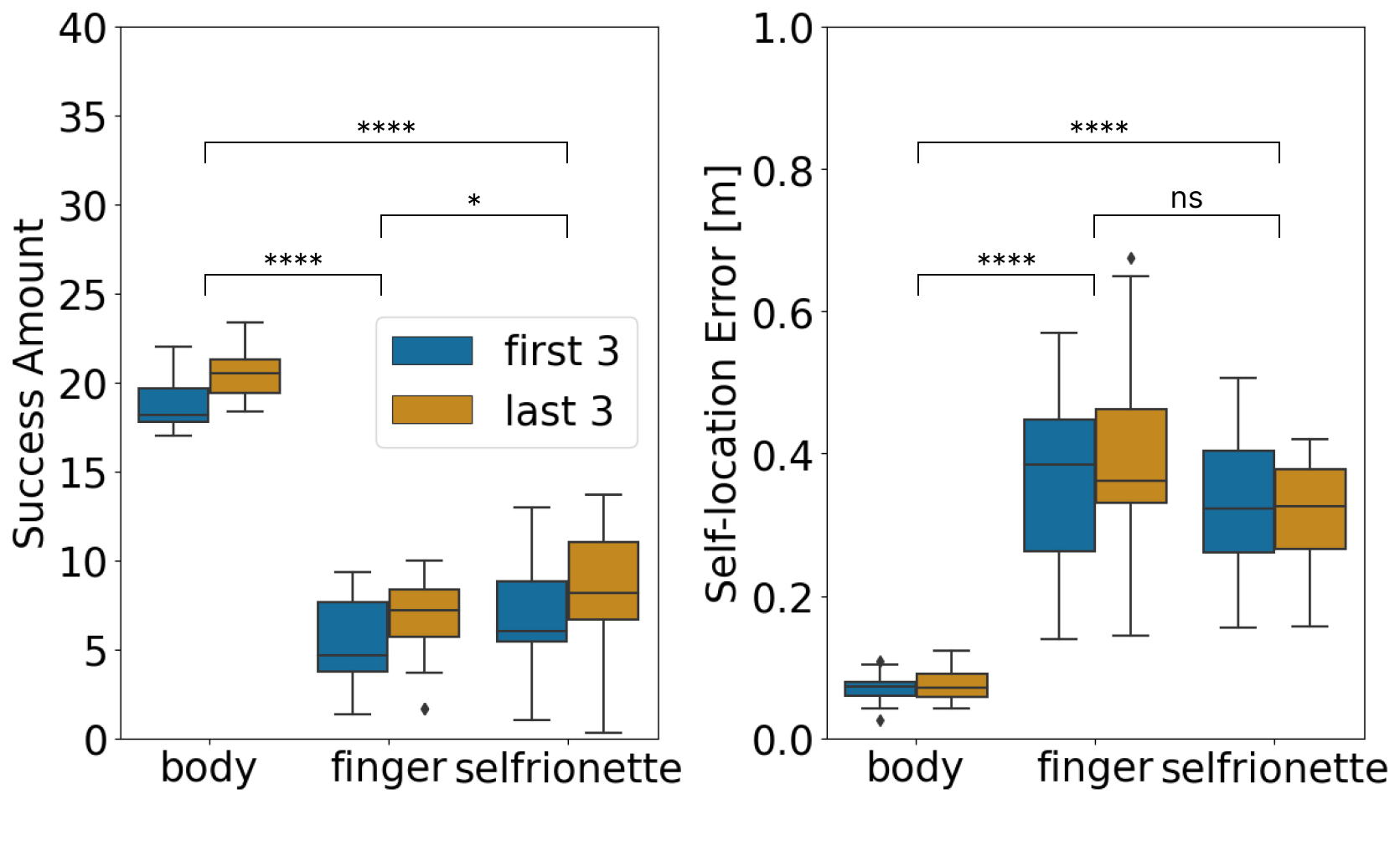}
    \caption{The average of task success amount (left) and self-location error (right) in fist three trials and last three trials}
    \Description{The average of task success amount (left) and self-location error (right) in fist three trials and last three trials}
    \label{fig:exp1time_result}
\end{figure}
\begin{figure*}
    \centering
    \includegraphics[width=\linewidth]{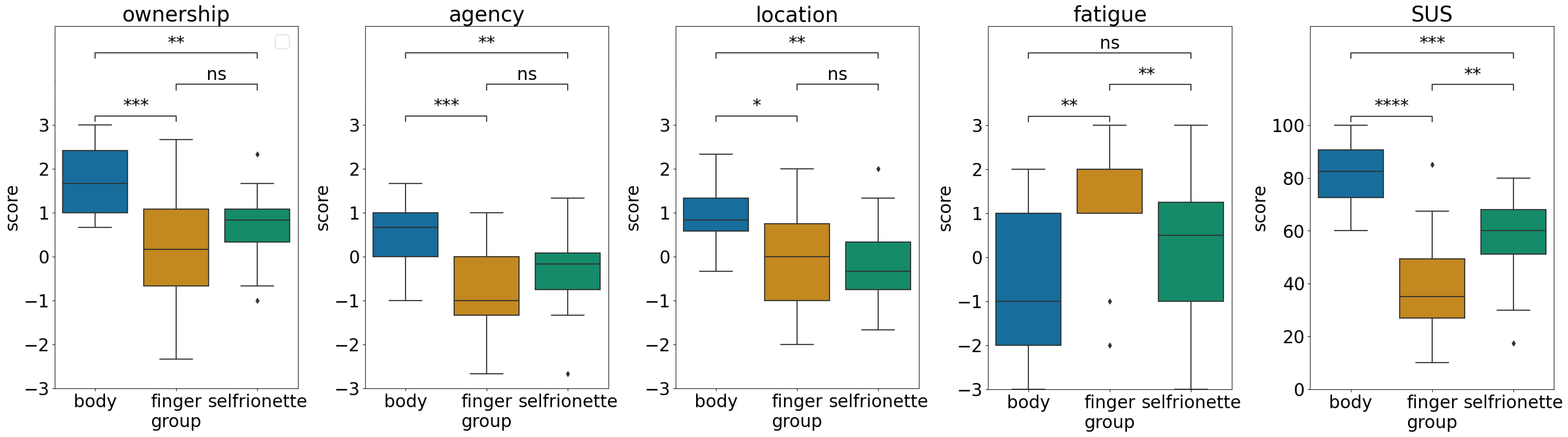}
    \caption{Result of Questionnaire}
    \Description{Result of Questionnaire}
    \label{fig:exp1_questionnaire}
\end{figure*}

Figure \ref{fig:exp1_task_time} shows the progression of reaching time for each trial and Figure \ref{fig:exp1time_result} (left) shows the corresponding average success amount of reaching in first 3 trials and last 3 trials. 
There is a clear trend indicating a decrease in reaching time as trials advance.
The convergence of learning curves suggests similar learning duration between the Finger and Selfrionette conditions.
A two-way ANOVA (\textit{manipulation condition} v.s. \textit{learning (first 3 and last 3)}) was used to analyze the results. 
When the normality assumption was violated (Shapiro-Wilk test), Aligned Rank Transform (ART) was used for the data. 
The results show a significant main effect in the manipulation condition factor ($F(2, 102)=111.5, p<.01$) and learning factor ($F(1, 102)=12.9, p<.01$). Then, post-hoc tests (Wilcoxon signed rank test with Shaffer's collection) showed significant differences between all pairs; body versus finger ($p<.01, r=1.1$), body versus Selfrionette ($p<.01, r=1.1$), and finger versus Selfrionette ($p=.047, r=.33$).

Figure \ref{fig:exp1time_result} (right) shows the average self-location task error in first 3 trials and last 3 trials. 
A two-way ANOVA (\textit{manipulation condition} v.s. \textit{learning (first 3 and last 3)}) was used to analyze the results and when the sphericity assumption was violated (Mauchly’s sphericity test), the degrees of freedom were adjusted using the Greenhouse-Geisser correction. Then, a significant main effect was found in the manipulation condition factor ($F(1.54, 26.14)=89.34, p<.01$). Post-hoc test (paired t-test with Shaffer's collection) showed that the self-location error of body condition was significantly lower than finger ($p<.01, r=.97$) and Selfrionette condition ($p<.01, r=.94$).

Figure \ref{fig:exp1_questionnaire} includes box plots illustrating the responses to each questionnaire.
An exact Wilcoxon signed-rank test with Shaffer's collection was applied to each of the five items: ownership, Agency, Location, Fatigue, and SUS. The results show significant differences in ownership ($p<.01, r=.77$), agency ($p<.01, r=.72$), and location ($p<.01, r=0.78$) score between body and Selfrionette, as well as ownership ($p<.01, r=.96$), agency ($p<.01, r=.95$), and location ($p=.03, r=.56$) between body and finger conditions. 
However, no significant differences were observed between Selfrionette and finger conditions.
Concerning fatigue, significant difference was found between body versus finger conditions ($p<.01, r=.83$) and finger versus Selfrionette conditions ($p<.01, r=.52$).
Moreover, significant differences were observed for SUS scores across all combinations;body vs Selfrionette ($p<.01, r=.86$), body vs finger ($p<.01, r=1.0$), and Selfrionette vs finger ($p<.01, r=.77$).

\subsection{Discussion}

The aim of this study was to investigate the extent to which Selfrionette could replicate avatar movements.

From the results of reaching time across conditions, it was found that Selfrionette and Finger could operate at approximately 47\% and 42\% of the speed of actual bodily movements, respectively. This also leads to the results that Selfrionette achieves more success amount than finger method.
While the Finger method lacked means for speed improvement due to one-to-one mapping between finger and arm mobility space, Selfrionette has some potential for speed enhancement through force-to-motion parameters adjustments in equation (\ref{eq:relation}).

Regarding embodiment questionnaire scores, the Body condition scored the highest, with no significant difference observed between Selfrionette and Finger conditions.
Participants found it challenging to achieve the same level of embodiment with Selfrionette or Finger as with the Body condition after a short experience of approximately 5 minutes. 
To enhance embodiment through these methods, longer exposure or multisensory experiences \cite{Botvinick1998-cy}, including hapitc interactions as categorized in section 4, may be necessary.

Additionally, we investigated whether self-location could be acquired through reaching learning, but after 10 trials, both Selfrionette and Finger were unable to achieve the same level of accuracy as Body condition. 
This suggests that at the beginning of the learning, visual feedback has a significant impact on avatar control with Selfrionette and Finger, and that approximately five minutes of learning did not lead to an update in the body schema. 
It is believed that longer training is necessary to estimate self-location based solely on proprioception information, as in the real world.
Furthermore, once the body schema is updated, there is potential for improvements in reaching task performance and bodily sensation. Therefore, validation of tasks for long-term learning and acquisition of new body schema is necessary.

Concerning fatigue, it was anticipated that Selfrionette and Finger conditions would result in lower fatigue scores than the Body condition because of its small effort for moving their avatar. 
However, the Finger condition yielded higher fatigue scores, likely due to the fingers being frequently placed in awkward positions during three-dimensional movements. 
Additionally, some participants reported feeling fatigued in the Selfrionette condition due to the tight grip required by the fingers, as indicated in their free responses.
Furthermore, the limited number of learning iterations in this experiment and the conservative settings of force input and movement gain to achieve accurate reaching movements may have contributed to finger fatigue.
To operate avatars with minimal force, dynamic gain learning may be necessary. 
This involves starting the learning process with low gain to allow users to acclimate to the movements and gradually increasing the gain over time.

System usability scores were highest for Body, followed by Selfrionette, and then Finger. 
The intuitive direct mapping of bodily movements in the Body condition likely contributed to this. 
For Selfrionette and Finger, mapping force exertion onto motion was more user-friendly than expanding finger mobility for avatar actions, especially given limited physical space for full-body control.

In summary, this user study suggests that for efficiently and accurately executing simple movements like reaching tasks, the one-to-one mapping of the Body condition is most suitable, followed by the self-rionette and then Finger conditions.
In the following user study, we plan to go beyond simple motor tasks and investigate the sense of embodiment towards avatars, mainly focusing on haptic interactions, a possible advantage of the Selfrionette.

\section{User Study 2}\label{sec:u2}
In User Study 1, the Body method excelled in simple motor tasks. 
However, our proposition of the Selfrionette system was based on the belief that it could offer a higher-quality haptic experience compared to other methods in haptic interaction. 
User Study 2 investigated which method provides a more realistic haptic experience and enhances the sense of embodiment with avatars across three different haptic interaction methods.

\subsection{Materials}
\subsubsection{Haptic properties}
\begin{figure}
    \centering
    \includegraphics[width=\linewidth]{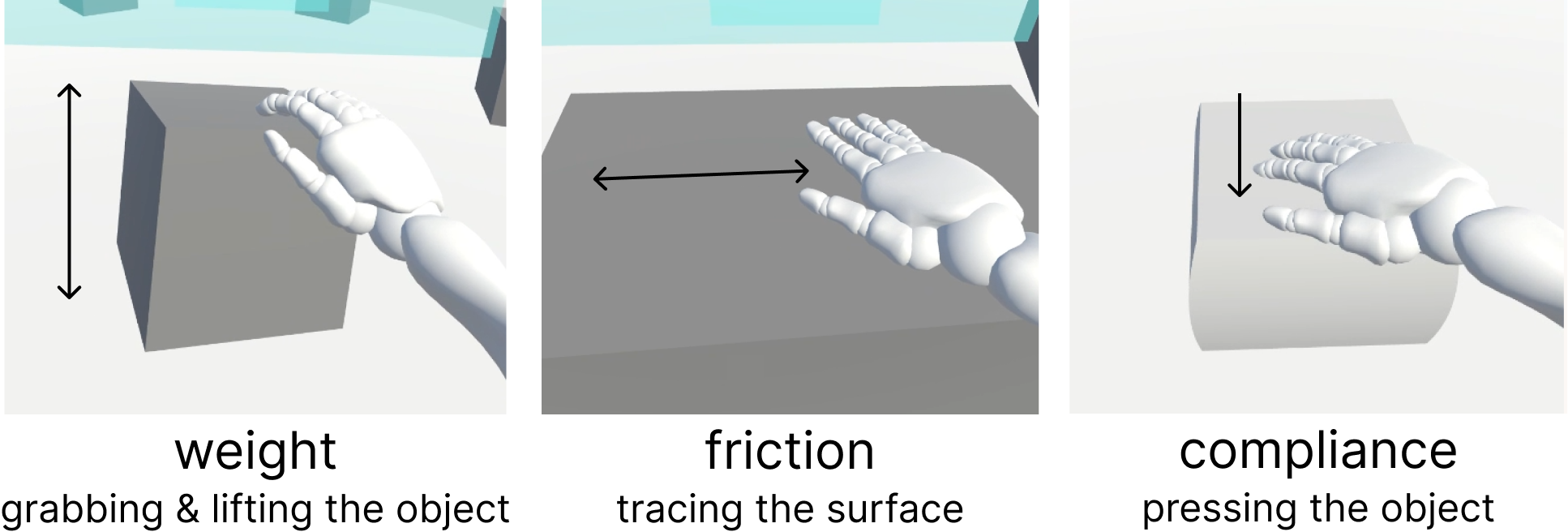}
    \caption{Interaction with objects of each haptic property.}
    \Description{Interaction with objects of each haptic property.}
    \label{fig:happrop}
\end{figure}

To achieve haptic interaction through avatars, we focused on the properties of materials of virtual object.
From the classification of Material properties in section \ref{sec:hap}, we implemented \textit{weight}, \textit{friction}, and \textit{compliance} properties to virtual objects.
Figure \ref{fig:happrop} illustrates the interaction with haptic objects through avatars.
The implementation of each Haptic property in Selfrionette aligns with the method outlined in section \ref{sec:hap}.
\revise{
For rendering weight, we applied equation (\ref{eq:weight}) to (\ref{eq:haptic_interaction}). 
Users used middle finger to grasp the virtual box.
For rendering friction, we applied equation (\ref{eq:friction}) to (\ref{eq:haptic_interaction}). 
For rendering compliance, we applied equation (\ref{eq:compliance}) to (\ref{eq:haptic_interaction}).
}
For rendering haptic feedback in the Body and Finger methods, we referenced prior works for each haptic property. 
Both methods use pseudo-haptics technique, where the avatar's movement changes upon interaction. 
Weight is represented by displacement changes when lifting an object, based on \cite{Lecuyer2009-cl, Samad2019-va}. 
Friction is depicted by displacement changes when tracing an object surface, based on \cite{Narumi2017-yk}. 
Compliance involves applying gain to displacement based on compression, expressed by the object's deformation \cite{Ban2014-uv}.

\subsection{Participants}
A total of 8 individuals participated, consisting of 6 males and 2 females, with an average age of 25.38 $\pm $ 2.77 years old. 
All participants were recruited through social media, unaware of the true purpose of the experiment.

\subsection{Haptic Properties Sorting Task}
To expose users to various haptic properties, we conducted a dummy task where they were asked to reorder the magnitude of haptic properties.
For each of weight, friction, and compliance, five degree of properties were provided. 
Users estimated the degree of properties by interacting with them through avatars and performed a ranking task.
A single object was displayed in a virtual space in front of the user, allowing interaction.
By pressing the left and right sides of the foot pedal, users could switch the interactable object. 
While interacting with different objects, they ranked them, and the final ranking was recorded.

Here, five distinct parameters were established for each haptic property by segmenting the range between a designated baseline and an upper limit into four equal intervals. 
The baseline was determined by the absence of pseudo-haptics techniques, representing the lower limit, while the upper limit was defined by the maximal parameter where pseudo-haptics techniques could be adapted, or the user interaction is just barely possible. 

\noindent {Weight Property:}
\begin{itemize}
    \item \textit{Lower Limit:} No gain applied under body and finger conditions (motion gain of 1); virtual object weight of \SI{0}{kg} under Selfrionette.
    \item \textit{Upper Limit:} Unliftably heavy state under body and finger conditions (motion gain of 0); virtual mass of \SI{1.6}{kg} under Selfrionette.
\end{itemize}

\noindent{Stiffness Property:}
\begin{itemize}
    \item \textit{Lower Limit:} No gain applied under the body and finger conditions (motion gain of 1); spring coefficient of 0 under Selfrionette.
    \item \textit{Upper Limit:} Virtual object does not deform (motion gain of 0 under the body and finger conditions; positive infinity spring coefficient under Selfrionette). The second stiffest object under Selfrionette has a spring coefficient of \SI{300}{N/m} with a damping coefficient of \SI{10}{Ns/m} for stability.
\end{itemize}

\noindent{Friction Property:}
\begin{itemize}
    \item \textit{Lower Limit:} No gain applied under the body and finger conditions (motion gain of 1); static and dynamic friction coefficients of 0 in Unity's physic material\footnote{https://docs.unity3d.com/2022.3/Documentation/Manual/class-PhysicMaterial.html} under Selfrionette.
    \item \textit{Upper Limit:} Hand fails to follow surface tracing actions under the body and finger conditions (motion gain of 0); maximum static and dynamic friction coefficients of 1 in Unity's physic material under Selfrionette.
\end{itemize}

\subsection{Questionnaire}
A questionnaire was administered to evaluate participants' subjective experiences regarding sense of embodiment, haptic experience and mental workload.
For sense of embodiment, we utilized the same questionnaire as in User Study 1.
Regarding haptic experiences, we prepared a question: one regarding the realism of the experience: ``The sensations of (weight, friction, compliance) presented in the virtual space closely resembled those I perceive in the real world.''
Each response was scored on a seven-point Likert scale (-3 = strongly disagree; 3 = strongly agree).
Additionally, mental workload was assessed using the NASA-TLX scale \cite{Hart1988-nt}.

\subsection{Procedure}
The experiment comprised three main components: the haptic properties sorting task and the questionnaire.
Participants engaged in these activities while wearing the HMD. 
The entire procedure lasted approximately 1 hour.
The participants
\begin{enumerate}
\renewcommand{\labelenumi}{\arabic{enumi})}
    \item read and signed an experimental consent form.
    \item received instructions and an overview of the task.
    \item completed haptic properties sorting task in weight, friction and compliance condition.
    \item completed the questionnaire.
    \item repeat (3) and (4) in each avatar control condition. 
    \item completed the post questionnaire.
\end{enumerate}

\subsection{Result}
\begin{figure}
    \centering
    \includegraphics[width=\linewidth]{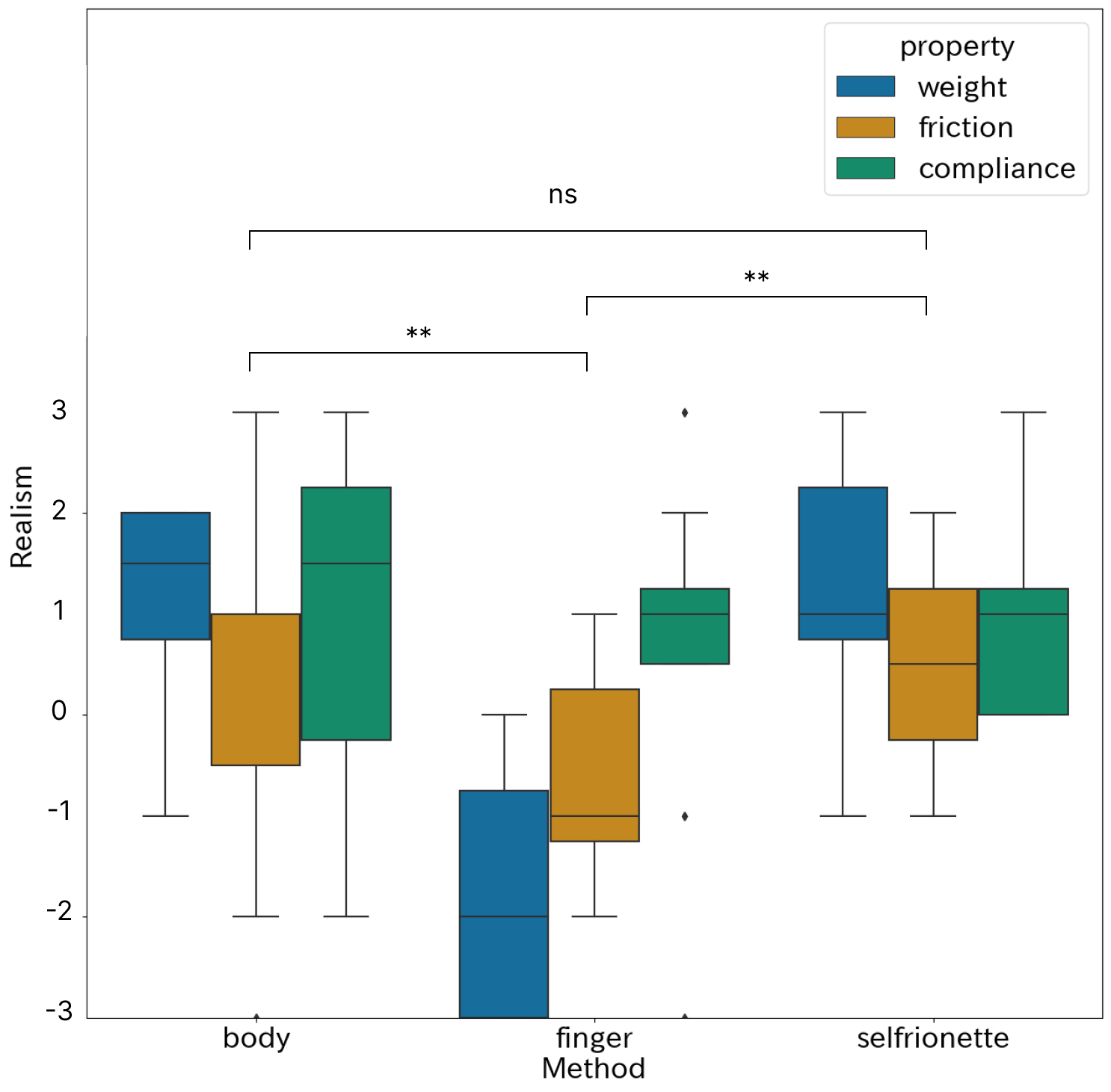}
    \caption{Result of realism score of haptic interaction.}
    \Description{Result of realism score of haptic interaction.}
    \label{fig:exp2_happrop}
\end{figure}

\begin{figure*}
    \centering
    \includegraphics[width=0.8\linewidth]{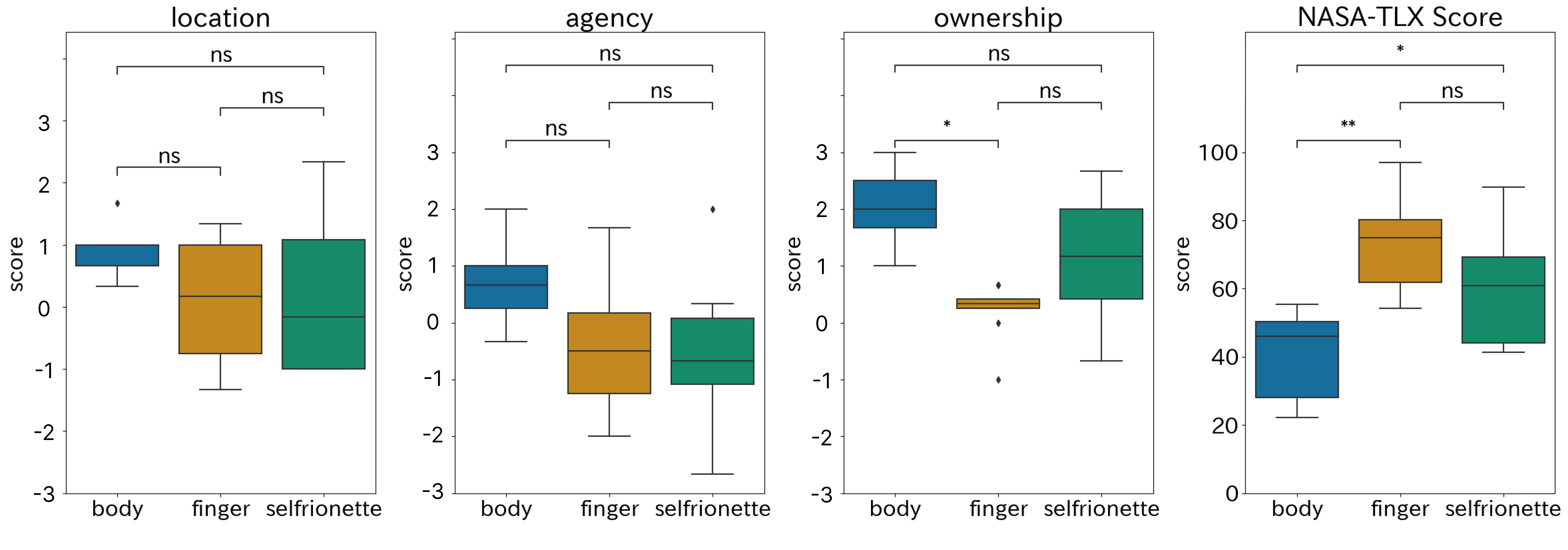}
    \caption{Result of sense of embodiment and NASA-TLX Score.}
    \Description{Result of sense of embodiment and NASA-TLX Score.}
    \label{fig:exp2_soe}
\end{figure*}

Figure \ref{fig:exp2_happrop} shows box plots of the realism score of each haptic interaction.
A two-way ANOVA (\textit{manipulation condition} v.s. \textit{haptic property}) was used to analyze the results. When the normality assumption was violated (Shapiro-Wilk test), ART was used for the data. The results show a significant main effect in the manipulation condition factor ($F(2, 63)=7.0, p<.01$). Then, post-hoc tests (Wilcoxon signed rank test with Shaffer's collection) showed that the realism score of body condition is significantly higher than that of finger condition ($p<.01, r=.58$) and that of Selfrionette condition is significantly higher than that of finger condition ($p<.01, r=.62$).

Figure \ref{fig:exp2_soe} includes box plots of the responses to embodiment questionnaire and the NASA-TLX score.
Wilcoxon signed rank test for each embodiment score (ownership, agency and location) and paired t-test for NASA-TLX score with Shaffer's collection were conducted. Consequently, significant differences were found between the ownership score of body condition versus that of finger condition ($p<.01, r=.94$). Moreover, the results show that the NASA-TLX score of body condition was significantly lower than that of finger condition ($p<.01, r=.86$) and that of Selfrionette condition ($p=.049, r=.67$). \revise{
Each element of the NASA-TLX (mental demand, physical demand, temporal demand, performance, effort, frustration) was individually analyzed.
The analysis revealed that the trends in scores for mental demand, physical demand, effort, and frustration closely mirrored the overall score trends. Conversely, the scores for temporal demand and performance were consistently low.
}

\subsection{Discussion}
Initially, it was anticipated that force-input methods would offer a more realistic haptic experience compared to motion-based techniques.
However, the outcomes of haptic realism score from both body tracking and Selfrionette were nearly identical. 
This may suggest that the discrepancy between the user's perceived body position and the avatar's position, managed through pseudo-haptics in Body condition, was minimal enough not to disrupt the user experience.
Moreover, the complexity of operating Selfrionette hindered users from executing desired movements, potentially limiting their ability to engage fully with the haptic properties. 
Further investigation is needed to ascertain whether proficient Selfrionette users perceive a heightened sense of haptic realism.
On the other hand, Selfrionette performed better in providing a realistic haptic experience compared to the finger method. 
For users trying out both methods for the first time without a fully developed understanding of how they work, Selfrionette provided sensations more akin to real haptics.
This is because Selfrionette allows users to perceive the force applied by their fingertips, enhancing the sense of haptic realism.

Regarding Embodiment scores, no significant difference was observed between Selfrionette and Body. 
In User Study 1, which employed a simple motor task, there was a discrepancy in Embodiment scores between Selfrionette and Body. 
However, this difference was not observed in the current task. 
This could be attributed to the inclusion of haptic information, absent in User Study 1, which may have accelerated the sense of embodiment towards the avatar. 
While further validation with an increased number of participants is necessary, it has been realized that incorporating appropriate haptic interactions when using Selfrionette could enhance embodiment towards the avatar.

Furthermore, in terms of NASA-TLX scores, Body exhibited better scores compared to Selfrionette and Finger.
As mentioned earlier, this could be attributed to the fact that both Selfrionette and Finger were novel manipulation methods for users, leading to an increased mental workload during operation.
Particularly, comments suggested that manipulating objects while moving the body, as required in the Weight condition, was considerably more challenging with Selfrionette and Finger compared to Body.
To make these novel methods more intuitive, improvements in hardware and mapping strategies are needed. Further details will be discussed in the Future Work section.

In conclusion, while Selfrionette imposes a higher mental workload compared to body during operation, its haptic experience quality is equivalent.
Moreover, when compared to the Finger method, Selfrionette offers a higher quality haptic experience, allowing users to feel embodiment in their avatars and enabling a realistic haptic experience within limited space.

\section{Limitation and Future Study}
\subsection{Learning and Gain Optimization}
\begin{figure}
    \centering
    \includegraphics[width=0.8\linewidth]{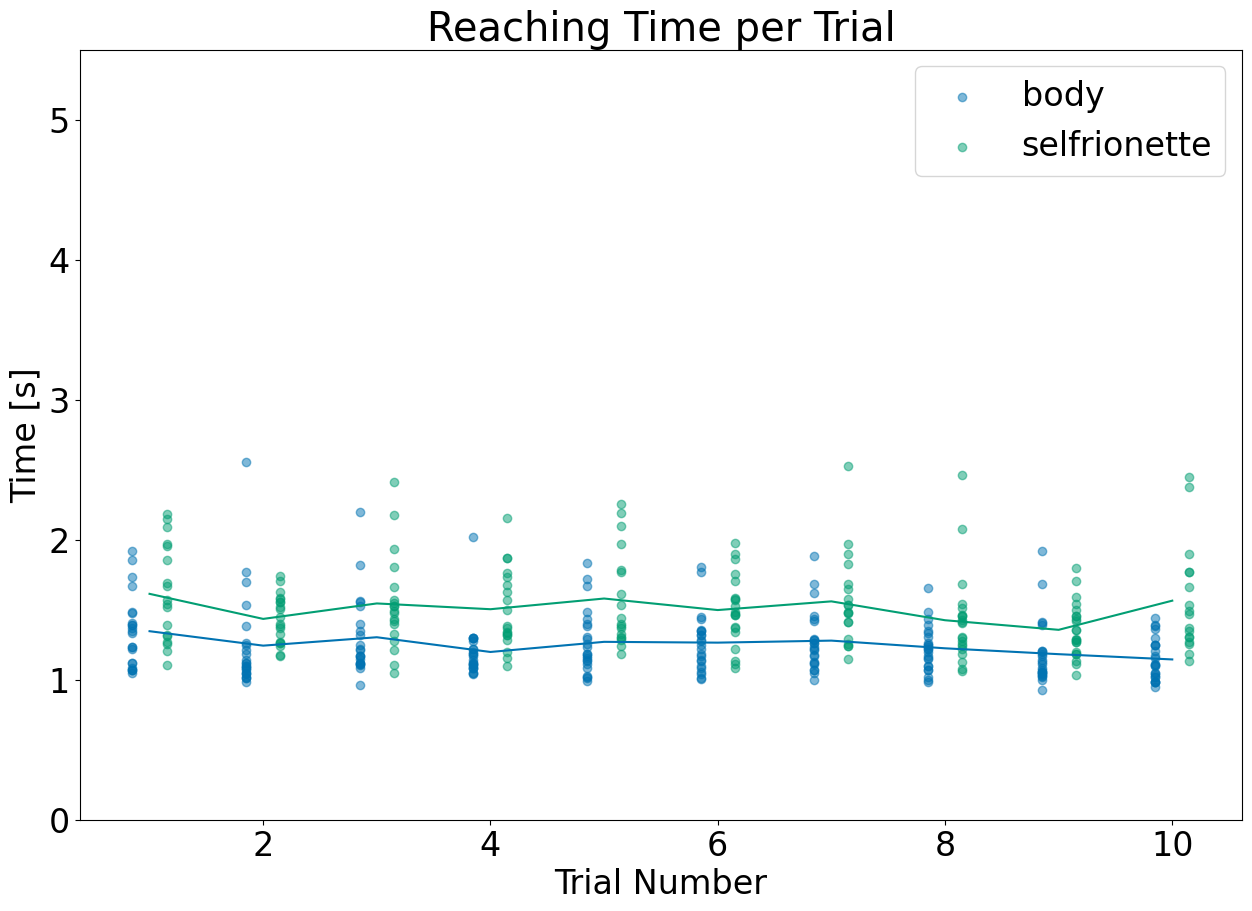}
    \caption{Task execution time per trial by experienced users on the same task as User study 1}
    \Description{Task execution time per trial by experienced users on the same task as User study 1}
    \label{fig:results_experienced}
\end{figure}

In the Selfrionette system, users re-associate body movements through finger force input.
User Study 1 showed that the operation is intuitive, with users typically adapting after performing the task for about 30 seconds, repeated almost five times.
However, a longer usage duration is needed to instill a sense of embodiment in the avatar. Some participants struggled even after 10 repetitions. 
Despite this, an experienced user (one of the authors) showed that with practice, performance in the Selfrionette condition could reach about 83\% of the body condition scores (Body average time: 1.24 s, Selfrionette average time: 1.50 s) (Figure \ref{fig:results_experienced}), suggesting that updating the body schema using Selfrionette is feasible.

Adjusting the force and motion gain could potentially shorten the initial learning process. 
In our user study, we used conservative low force gain values to ensure ease of avatar control for most participants. 
However, the gain settings can impact the operation's accuracy, comfort, and energy efficiency. 
Future studies should explore dynamic gain tuning techniques, gradually increasing the gain as users become more familiar with the operation, allowing quicker and more precise movements with less force.

\revise{
\subsection{Full-Body Interaction}
To evaluate the effectiveness of the proposed method in experiments, this paper focuses on hand interaction as a first step, although we implemented walking interactions with using thumbs on trial. 
In future research, it will be crucial to experimentally verify the effectiveness of Selfrionette for the entire body, including the legs.
Preliminary user studies on the WIP method implemented with the thumbs in this study indicate that the current method of walking control is not easy to master and requires longer practice compared to arm control using the index finger. 
Therefore, future research should focus on developing a more intuitive leg operation mapping and walking control system that facilitates easy and rapid learning.
It has been established that using alternating index trigger inputs as a locomotion technique in VR can alleviate motion sickness compared to joystick locomotion\cite{sarupuri2017triggerwalking}. Therefore, it is hypothesized that Selfrionette WIP, which utilizes alternating force inputs, may also reduce motion sickness while providing a sensation of real walking. This hypothesis should be investigated in future research.
Furthermore, while the current implementation achieves walking interactions using 1 DoF input from the thumbs, we plan to implement 3 DoF control for the legs, similar to the arms, to enable more complex leg interactions.
Moreover, it will be necessary to investigate the full-body manipulation using multiple fingers from a view of cognitive load.
}

\subsection{Ergonomic Design for Efficient Force Input}
To allow users to position their hands naturally, the Selfrionette hardware uses a mouse-like casing. 
However, this hand posture made it difficult to apply force in certain directions with the index finger. Users struggled to exert force in the direction of index finger spreading and had difficulty moving it forward and backward, sometimes resulting in excessive force and unstable avatar operation.
Redesigning the device to allow easier force application in all three axes, especially with the index finger, such as a semi-gripped position, could improve comfort and operability.

\section{Conclusion}
The Selfrionette system is an interface that helps people move their bodies in virtual reality (VR) spaces and improves haptic interactions.
It is especially useful for people who have limited space or physical disabilities, allowing them to fully enjoy immersive VR experiences.
With Selfrionette, interactions involving the entire body, such as reaching for objects, grasping them, and walking, can be controlled solely by the force applied by the fingers. 
Furthermore, diverse haptic properties can be imparted to these interactions.

In summary, it has been established that body tracking methods excel in intuitiveness and are well-suited for completing simple motor tasks. 
On the other hand, it was observed that there is a difference in the ability to perform simple motor tasks between Selfrionette and Finger conditions, both of which allow for full-body operation in limited space.
Additionally, methods utilizing Selfrionette excel in system usability and haptic experience quality compared to finger-motion methods, and they can provide realistic haptic feedback comparable to body tracking methods.

In conclusion, Selfrionette is a promising tool for enriching haptic interactions in VR spaces, especially beneficial for those who value limited spaces or who cannot move their arms and legs as much as they would like in the real world.
Also, depending on how and for how long the user uses the Selfrionette, it may exceed body tracking in terms of motion velocity, haptic experience, and embodiment factors.

\begin{acks}
This work was partially supported by JST ACT-X (JPMJAX22KD), JST Moonshot Research \& Development Program (JPMJMS2013), Grant-in-Aid for Research Activity Start-up (23K20007), and Grant-in-Aid for Early-Career Scientists (24K20818).
\end{acks}

\bibliographystyle{ACM-Reference-Format}
\bibliography{ref}

\end{document}